\documentclass[conference, 10pt]{IEEEtran}
\IEEEoverridecommandlockouts 
\usepackage[]{graphicx}
\usepackage{caption}
\usepackage{subfig} 
\usepackage{amsmath}
\usepackage{amssymb}
\usepackage{epsfig}
\usepackage{cite}
\usepackage{color}
\usepackage{balance}
\usepackage{amsmath}
\usepackage{amsfonts} 
\usepackage{graphicx}
\usepackage{pdfpages}
\usepackage[T1]{fontenc} 
\usepackage{array}
\usepackage{url}

\usepackage{color}
\usepackage{wrapfig}
\usepackage{lipsum}
\usepackage[hidelinks]{hyperref}
\usepackage{multirow}
\usepackage{tabularx}
\usepackage{tikz}
\usepackage{nth} 
\usepackage{algpseudocode} 
\usepackage{algorithm,algpseudocode}
\usepackage{breqn}
\usepackage{todonotes}
\usepackage{verbatim}

\begin{document}
\title{An Interoperable Open Data Portal\\ for Climate Analysis}
\author{\IEEEauthorblockN{
Jiantao~Wu\IEEEauthorrefmark{1}\IEEEauthorrefmark{2},
Huan~Chen\IEEEauthorrefmark{2},
Fabrizio~Orlandi\IEEEauthorrefmark{1}\IEEEauthorrefmark{3},
Yee Hui Lee\IEEEauthorrefmark{4},
Declan O'Sullivan\IEEEauthorrefmark{1}\IEEEauthorrefmark{3}, and
Soumyabrata Dev\IEEEauthorrefmark{1}\IEEEauthorrefmark{2}
}
\IEEEauthorblockA{\IEEEauthorrefmark{1} ADAPT SFI Research Centre, Dublin, Ireland}
\IEEEauthorblockA{\IEEEauthorrefmark{2} School of Computer Science, University College Dublin, Ireland}
\IEEEauthorblockA{\IEEEauthorrefmark{3} School of Computer Science and Statistics, Trinity College Dublin, Ireland}
\IEEEauthorblockA{\IEEEauthorrefmark{4} School of Electrical and Electronic Engineering, Nanyang Technological University (NTU), Singapore}
\thanks{This research was partially funded by the EU H2020 research and innovation programme under the Marie Skłodowska-Curie Grant Agreement No.~713567 at the ADAPT SFI Research Centre at Trinity College Dublin. The ADAPT Centre for Digital Content Technology is funded under the SFI Research Centres Programme (Grant 13/RC/2106_P2) and is co-funded under the European Regional Development Fund.}

\vspace{-0.6cm}
}

\maketitle

\begin{abstract}
This work proposes an open interoperable data portal that offers access to a Web-wide climate domain knowledge graph created for Ireland and England’s NOAA climate daily data. There are three main components contributing to this data portal: the first is the upper layer schema of the knowledge graph --the climate analysis (CA) ontology -- the second is an ad hoc SPARQL server by which to store the graph data and provide public Web access, the last is a dereferencing engine deployed to resolve URIs for entity information. Our knowledge graph form of NOAA climate data facilitates the supply of semantic climate information to researchers and offers a variety of semantic applications that can be built on top of it.
\end{abstract}

\IEEEpeerreviewmaketitle

\section{Introduction}
Nowadays, with challenges related to energy shortages, global warming, air pollution \textit{etc.}, climate-related studies have become essential in academia and industry. Some researchers tend to use data-driven methods (\textit{e.g.} data modeling) to understand the climatic problems~\cite{manandhar2019data}. Those researchers usually need a great amount of data to support. Climate data centres such as KNMI Climate Explorer and NOAA publish tabular data (\textit{e.g.} RDB, CSV, JSON). However, raw tabular data sometimes can fail to express information of interest. For example, tabular data is hard to present the relationship between data. For another, many climate problems relate to a number of parameters beyond a single data source such as the CO$_2$ emission is relevant to air temperature, fetching and integrating data of different domains will be time-consuming for researchers. Recently, employing knowledge graph to store data has been popularly explored, which can benefit to solve these dataset-related problems. In this work, we create a knowledge graph portal\footnote{\url{http://jresearch.ucd.ie/linkclimate/}} to store Ireland’s and England’s NOAA climate daily summary data so that these climate data can easily benefit from semantic techniques.
\section{Background}
In this section, we give the background information about the original data source and some core techniques underpinning the knowledge graph.
\subsection{NOAA Climate Data}
National Centers for Environmental Information(NCEI) is one of the data centres of the National Oceanic and Atmospheric Administration (NOAA). NOAA's climate data has been widely used in different fields. It provides global datasets related to oceanic, atmospheric, and geophysical data. CSV, SSV, JSON, PDF and NetCDF format data can be accessed by RESTful APIs offered by NOAA. Our research transforms the CSV formatted climate data into RDF representation of the data so that it can be made available as a knowledge graph accessible as Linked Open Data through an accessible query endpoint.

\subsection{Ontology in Short}
Ontology represents the object's structures, attributes and relations~\cite{orlandi2019interlinking}.
It is the schema of the knowledge graph, grouping data into hierarchical classes and through pre-defined properties establishing connections between the instances of the classes~\cite{ehrlinger2016towards, singhal2012introducing}. It defines the relationships of concepts and conceptual properties. Ontology can help facilitate knowledge share and shared ontology can improve the reusability of knowledge~\cite{chandrasekaran1999ontologies}. 
\subsection{Linked Data}
Another fundamental used in the ontology model is the Linked Data. Linked Data stands for a principled approach to publishing and connecting structured data on the web~\cite{bizer2011linked}. It is well-performed in organizing and integrating data on the Web, improving information retrieval in the big data background. It provides a way to allow data from different sources to be connected and queried. The linked data can be queried with a semantic query language--SPARQL so that Resource Description Framework (RDF) format data associated with the knowledge graph can be searched and operated.
\section{Linked Climate}
In this section, we present the details of our proposed knowledge graph model on NOAA climate data. It is composed of a SPARQL server with data periodically updated from NOAA Climate Data Online and a dereferencing engine powered by LodView\footnote{\url{https://github.com/LodLive/LodView}}. The link climate knowledge graph data is structured by climate analysis(CA) ontology. We begin by introducing the climate analysis ontology, followed by a climate data query. We then describe climate data graph visualization at last.

\subsection{Climate Analysis Ontology}
In this project, we create a the CA ontology to fit the NOAA daily summary datasets~\cite{wu2021ontology}. This ontology consists of two main classes: 1) Station and 2) Observation. An example is shown in Figure 1. A station indicates the platform of climate observation that is spotted by the longitude and latitude geographical features and a literal name as to its attributes. An observation represents the climate attributes observed such as " temperature" and "precipitation". Observation usually connects with the associated station and it also connects with attributes.

\begin{figure}
    \centering
    \includegraphics[width=9cm]{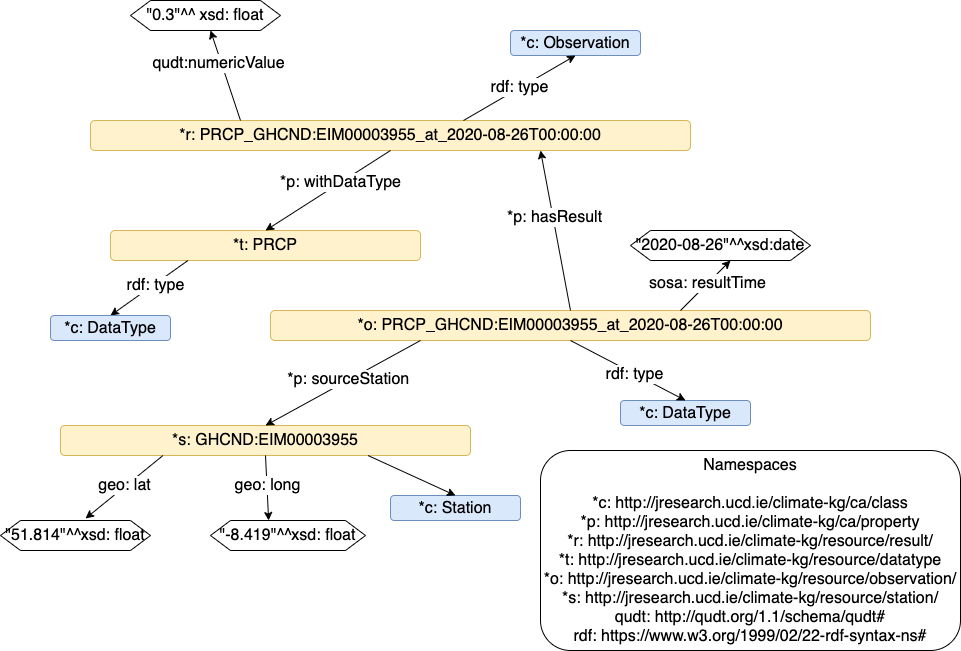}
    \caption{We model the climate knowledge graph model on NOAA climate data.}
    \label{fig:my_label}
\end{figure}

\subsection{Climate Data Query}
We use Jena Fuseki\footnote{\url{https://jena.apache.org/documentation/fuseki2/}} as the linked data publishing platform. Based on the CA ontology, we transform NOAA climate data to RDF triples to create data instances. The process implements through a python script, which runs weekly to send HTTP requests to NOAA to obtain the latest one-month observations both in Ireland and England.  

With the predesigned queries, users can readily get some actual climate data by defining a graph pattern in SPARQL to select out them. In a simplified SPARQL query, the query starts with key words “SELECT” and “WHERE”. The variables next to "SELECT" (starting with ?) are the variables selected for displaying their bindings (actual data) in the query results. The "\{...\}" after "WHERE" is where the graph pattern is placed.
\subsection{Climate Data Graph Visualization}
In this work, every data entity is encoded with a unique URI in the Fuseki triplestore. The data is dereferenced by virtue of the LodView tool. To visualize the data, users can query the SPARQL endpoint integrated in the data portal and click on the URI-encoded results to obtain detailed information about a specific entity. As we have configured the LodLive (see“view on Lodlive” on the LodView profile) as part of LodView, the results can also been visualized as node-extensible graphs which may assist users in clicking to continuously bring in new information, instead of refreshing to a new entity LodView profile page. An example is shown in Fig. 2.

\begin{figure}
    \centering
    \includegraphics[width=8cm]{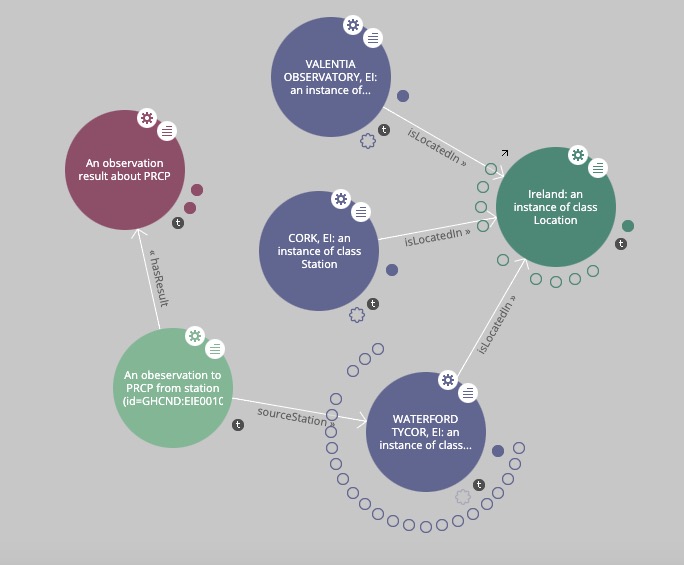}
    \caption{We provide the graphical representation of dereferencing navigation.}
    \label{fig:my_label}
\end{figure}

\section{Conclusion \& Future Work}
In this paper, we present our climate knowledge graph based on  Ireland’s and England’s NOAA climate daily summary datasets. It consists of a climate analysis ontology that is the schema of our climate knowledge graph; a SPARQL server that periodically updates data from NOAA and provides the place to readily get climate data with SPARQL; a LodView dereferencing engine that offers Climate Data entity visualization. In the future, we will make some further improvements to our model to enhance usability for climate researchers.


\bibliographystyle{IEEEtran.bst}

\end{document}